\documentclass[aps,prl,twocolumn,superscriptaddress,showpacs]{revtex4}
\usepackage{amsmath}
\usepackage{dcolumn}
\usepackage{graphicx}
\usepackage{latexsym}
\usepackage{float}

\begin{document}

\title{Origin of the exciton mass in the frustrated Mott insulator Na$_2$IrO$_3$}

\author{Zhanybek Alpichshev}
\affiliation{Department of Physics, Massachusetts Institute of Technology, Cambridge, Massachusetts 02139}

\author{Edbert J. Sie}
\affiliation{Department of Physics, Massachusetts Institute of Technology, Cambridge, Massachusetts 02139}

\author{Fahad Mahmood}
\affiliation{Department of Physics, Massachusetts Institute of Technology, Cambridge, Massachusetts 02139}

\author{Gang Cao}
\affiliation{Department of Physics and Astronomy, University of Kentucky, Lexington, Kentucky 40506}

\author{Nuh Gedik}
\email{gedik@mit.edu}
\affiliation{Department of Physics, Massachusetts Institute of Technology, Cambridge, Massachusetts 02139}

\date{\today}

\begin{abstract}
We use a three-pulse ultrafast optical spectroscopy to study the relaxation processes in a frustrated Mott insulator Na$_2$IrO$_3$. By being able to independently produce the out-of-equilibrium bound states (excitons) of doublons and holons with the first pulse and suppress the underlying antiferromagnetic order with the second one, we were able to elucidate the relaxation mechanism of quasiparticles in this system. By observing the difference in the exciton dynamics in the magnetically ordered and disordered phases we found that the mass of this quasiparticle is mostly determined by its interaction with the surrounding spins.
\end{abstract}

\pacs{75.10.Kt, 71.10.Li, 78.47.jj}

\maketitle    

The notion that any generic interacting many-body system near equilibrium can be described by a number of noninteracting excitations dubbed quasiparticles lies at the heart of modern condensed-matter physics \cite{pines}. This approach is extremely powerful and can be used to describe almost any many-body system known to date. However the exact character of the resulting quasiparticles can be very different from the properties of the original electrons and lattice ions. A notable example is the problem of a doped Mott insulator \cite{imada, lee}. Here an additional hole (or electron) cannot be thought of as a simple Bloch wave since, while propagating, it inevitably scrambles the surrounding magnetic order \cite{lee, brinkman}. The result is the so-called ``separation of spin and charge degrees of freedom'' in the original holes and electrons \cite{fradkin}. The charge is carried away by spinless quasiparticles called ``holons'' (positively charged) and ``doublons''(negative) and the spin by neutral ``spinons'' \cite{rokhsar}. In addition, strong correlations also affect the mass of a holon (doublon) making it much heavier as compared to a bare hole (extra electron). Intuitively this happens because in order for a holon or doublon to hop to the next lattice site it needs to wait for the spins to recover after the previous hop (because the holon/doublon is a quasiparticle) \cite{khomski, brinkman, shraiman, trugman, varma, kane}. The waiting time is determined by spin-spin interactions which are typically much weaker than orbital interactions, therefore the effective mass of holons and doublons becomes much larger compared to bare electron mass\cite{lee}.

There is strong experimental evidence that spin-charge separation takes place in actual materials \cite{kim}. On the other hand it is less clear if the correlations in Mott insulators indeed renormalize the quasiparticle mass. The challenge here is that although conventional equilibrium techniques can  observe enhanced carrier mass in materials known to be strongly correlated \cite{thomas}, being based on linear response they can tell very little on the \textit{origin} of the observed mass enhancement. Analogously the value of the proton mass was known for a long time, however it took developing quantum chromodynamics (QCD) to understand its origin \cite{peskin, durr}. Despite the intuitive appeal, the considerations in the previous paragraph heavily rely on the ideas specific to the Mott insulating state. On the other hand the mass enhancement by itself can arise due to a variety of other unrelated reasons including the more conventional polaronic effects \cite{pekar} (which might also be relevant for cuprates \cite{misch1,misch2}) or even simple single-particle band effects \cite{kittel}. In order to establish that a particular mechanism is indeed responsible for the given equilibrium properties (such as the effective mass) one necessarily needs to go beyond static linear-response probes. One way is to study the behavior of the system away from equilibrium on appropriate timescales with an ability to control each relevant degree of freedom (charge, spin, lattice, etc.) individually on appropriate timescales.

\begin{figure}[t]
\includegraphics[width=0.90\columnwidth]{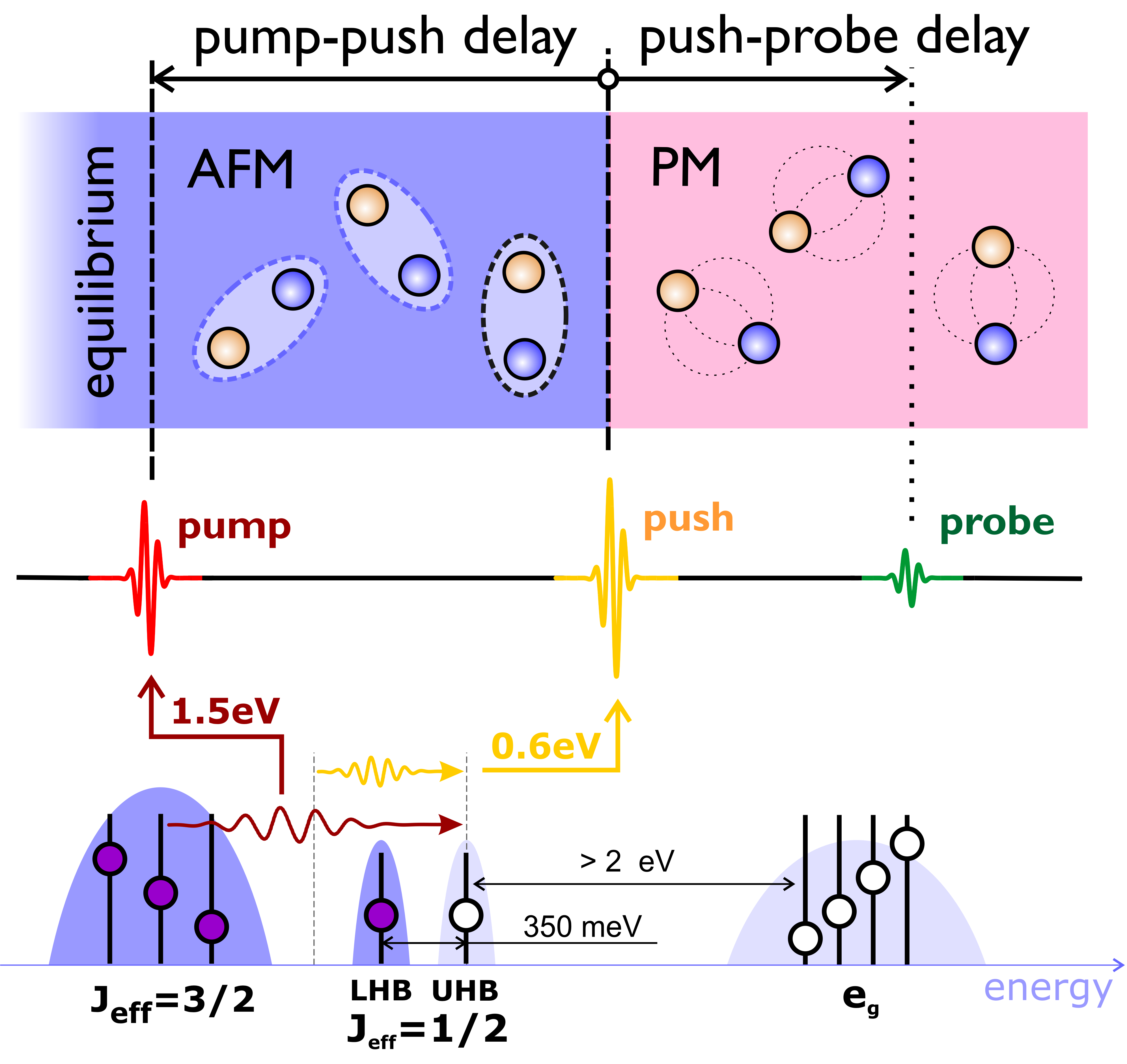}
\caption{A schematic illustration of the experiment. Top: a cartoon of the three pulse (``pump-push-probe'') process: doublons and holons are first created in an ordered [antiferromagnetic (AFM)] state with the pump pulse. Then order is melted and new disordered [paramagnetic (PM)] state is created by the push pulse arriving after some controllable time delay after the pump pulse. Doublons and holons form bound states in both AFM and PM, however in PM they are held together by Coulomb attraction only while in AFM there is a spin string connecting them which is responsible for enhanced mass of the bound state in AFM. Bottom: sketch of the band structure of Na$_2$IrO$_3$ \cite{damascelli} comparing it with the energy of the pump ($\hbar \omega = 1.5$ eV) and the push ($\hbar \omega =590$ meV) pulse energies.}
\label{hopping}
\end{figure}

\begin{figure*}[t]
\includegraphics[width=0.95\textwidth]{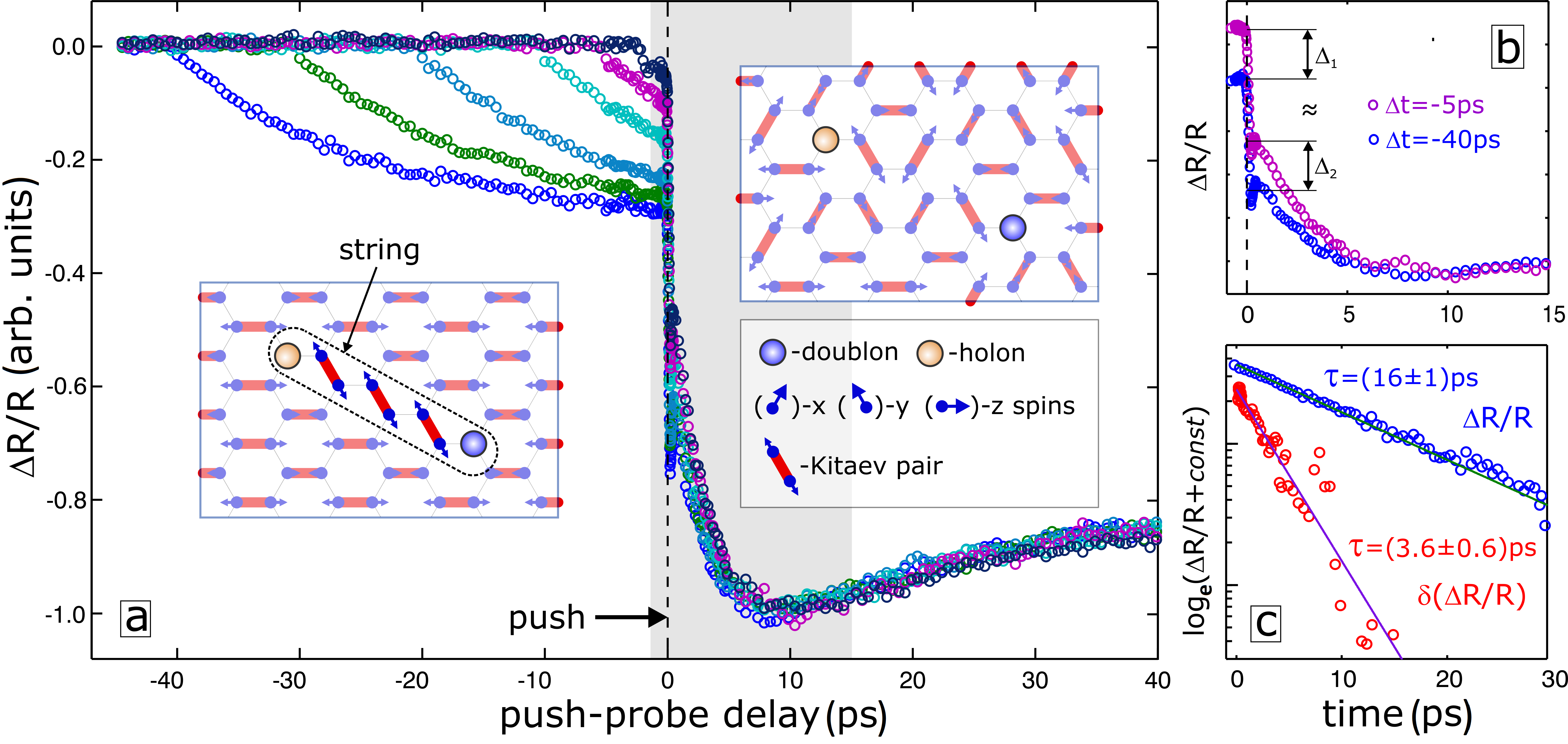}
\caption{a) Three-pulse traces with a strong push pulse ($\Phi_{\text{push}}=500$ $\mu$J/cm$^2$, $\Phi_{\text{pump}} = 2.5$ $\mu$J/cm$^2$) at $t=0ps$ taken at a base temperature of $T=5K$. The push pulse is heating the system above T$_N$ (see text); b) Zoom into region shaded in gray in (a) with only two traces with pump-push delay $\Delta t=-40$ ps (blue) and $\Delta t=-5$ ps (purple) shown for clarity. As can be seen the ``memory'' of the electronic system is not erased by the push pulse as the vertical difference proportional to the deviation from equilibrium is same before and after the push pulse. On the other hand this vertical difference decays much faster after the push pulse indicating that the excitons lose energy (not recombination) faster at $T>T_N$. This indicates that the mass of the Hubbard exciton in the disordered state [left inset to (a)] is smaller than that in the ordered state (right inset to (a)) due to the presence of a spin string (highlighted with a dashed line); c) the logarithm of the signal before the push pulse (blue) and of the difference between the signals with $\Delta t=-40$ ps and $\Delta t=-5$ ps delays between the pump and push pulses (purple) indicating that the relaxation process can be described as simple exponential decay in both cases justifying the simple relaxation picture adopted in the text.}
\label{aboveN}
\end{figure*}

In this paper we use time-resolved optical spectroscopy to determine the mechanism behind the quasiparticle mass renormalization in a frustrated Mott insulator. We study the behavior of the Hubbard exciton in Na$_2$IrO$_3$ which is a magnetically frustrated Mott system \cite{khaliullin1, khaliullin2, kitaev}. Previously it was found that at low temperatures the nonequilibrium charge excitations in it behave as doublons and holons \cite{alpichshev, hinton, nembrini} which can form bound states [``Hubbard excitons'' (HEs)] \cite{rixs}. In the magnetically disordered state these are more or less conventional excitons held together by Coulomb attraction. In contrast in the ordered low temperature phase (antiferromagnetic) the spins form a ``string'' between constituent doublons and holons \cite{alpichshev} (the string is a quasi-one-dimensional region of reorganized spin ordering that connects doublons and holons reflecting their fractional nature \cite{lee, fradkin, rokhsar}. See insets to Fig. \ref{aboveN}(a) and Ref. \cite{alpichshev}). By using a time-resolved technique developed for this work we can suppress the magnetic ordering at any stage of relaxation of HE and observe that the presence of the string slows down the relaxation dynamics of HE which signals an increase in its mass. This is expected if we accept that the spin string should also perturb the spin order as HE moves. Therefore we conclude that the mass of the Hubbard exciton is predominantly determined by the strong correlations between charge and spin degrees of freedom.

In a conventional pump-probe method the sample is excited by a short laser pulse called the pump and then a time delayed second pulse called the probe is sent to measure the nonequilibrium reflectivity of the sample. In this way it is possible to infer the details of the interactions within the system that determine its relaxation dynamics \cite{orenstein}. An upgraded time-resolved pump probe system used in this work features two separate laser pump pulses with appropriately chosen wavelengths. The first pulse (``pump'', $\hbar \omega = 1.55$ eV) is used to create excited doublons and holons which quickly form non-equilibrium HEs (but do not recombine during the course of experiment due to selection rules \cite{alpichshev, demler}). The second pumping pulse (``push'', $\hbar \omega = 0.6$ eV) is minimally coupled to electronic degrees of freedom as its energy is not sufficient to excite new electrons from the $J_{\text{eff}}=3/2$ band \cite{choi} (see Fig. \ref{hopping}) and the intraband excitations in the $J_{\text{eff}}=1/2$ bands are suppressed due to their narrow character ($W\ll \hbar \omega$) \cite{bjkim1}. The push pulse thus predominantly generates bosonic excitations and as such can be used to instantaneously destroy the magnetic order by melting it at any stage of HE relaxation. This can happen through a number of channels such as multiphonon near-infrared absorption \cite{rupprecht} or impulsive stimulated Raman scattering \cite{bloembergen, fleury, herring, raman, supp}.

For the double pump-probe (pump-push-probe) experiments we used an amplified laser system operating at the center wavelength of $790$ nm and the repetition rate of $30$ kHz whose output was used for optical parametric amplification (OPA) and white light supercontinuum (WLS) generation in a sapphire crystal to produce various pulses: $790$ nm ($1.55$ eV, fundamental) with a spot size $0.6$ mm full width at half maximum (FWHM) for the pump pulse; $2100$ nm ($0.6$ eV, OPA) with a spot size $250$ $\mu$m FWHM for the push; and $907$ nm ($1.38$ eV, WLS) with a spot size $150$ $\mu$m FWHM for probing, nondegenerate with pump to minimize noise coming from pump scattering. In all experiments reported in the main text the pump fluence and total power were chosen such that the measurements are performed in the low fluence regime, where the signal dynamics is independent of pump fluence \cite{alpichshev, supp}. Single crystals of Na$_2$IrO$_3$ were grown using a self-flux method from off-stoichiometric quantities of IrO$_2$ and Na$_2$CO$_3$. Similar technical details were described elsewhere \cite{sample1, sample2, sample3}. Samples were cleaved \textit{ex situ} before every measurement to expose fresh surface and placed under vacuum within a few minutes.  

Just as equilibrium optical conductivity data is used to interpret single pump-probe experiments we will interpret the double pump-probe (pump-push-probe) data presented in this paper relying on the analysis of the regular single pump-probe experiment on Na$_2$IrO$_3$ reported in \cite{alpichshev}. The summary of the relevant conclusions of \cite{alpichshev} is as follows: 1) The transient optical response of Na$_2$IrO$_3$ has a qualitatively different behavior below and above the ordering temperature $T_N=15$ K. In particular for temperatures $T<T_N$ the signal is independent of temperature and is a monotonous function of time that can be fit with a single exponential while for $T>T_N$ the transient signal is nonmonotonous with an extremum whose position is approaching the origin with increasing temperature. This indicates that the relaxation dynamics in Na$_2$IrO$_3$ is determined by magnetism as opposed to other possible channels such as electron-lattice interactions. 2) The slow low temperature signal is due entirely to bound states of doublons and holons [Hubbard excitons (HEs)] while the high temperature signal is a mixture of the response from HEs and the doublon-holon ``plasma.'' 

\begin{figure}[t]
\includegraphics[width=\columnwidth]{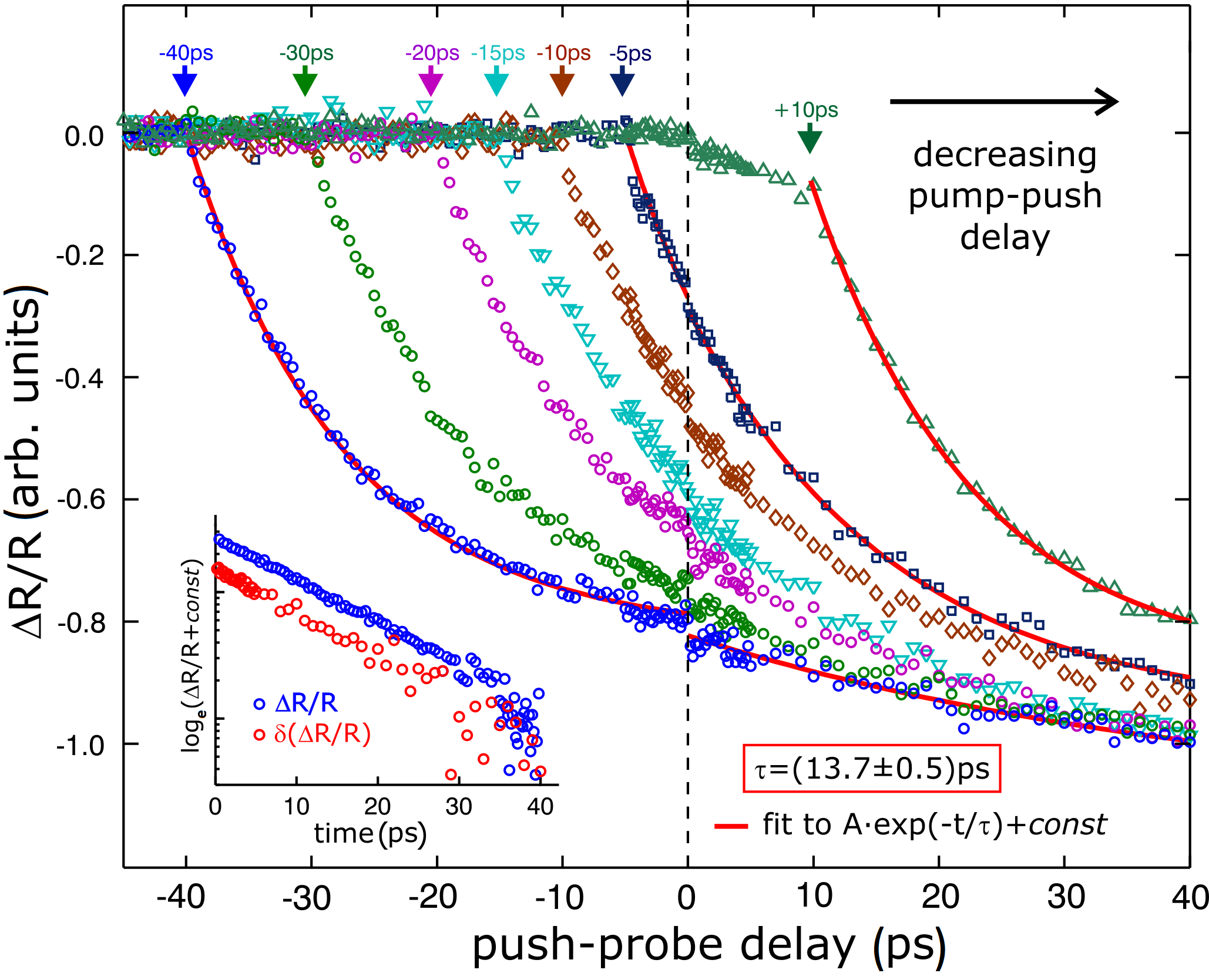}
\caption{Three pulse data with a weak push pulse insufficient to melt the magnetic order (see text) taken at $T=5$ K. $\Phi_{\text{push}}=100$ $\mu$J/cm$^2$, $\Phi_{\text{pump}}= 4$ $\mu$J/cm$^2$. The push pulse is at $0$ ps; solid red lines: fits to a single exponential decay with the same time constant everywhere. Inset: logarithm of the signal before push pulse (blue) and the logarithm of the difference between the lowest (pump push delay $\Delta t=-40$ ps) and highest ($\Delta t=-5$ ps) curves (red). As can be seen, a weak push pulse does not affect the relaxation dynamics.}
\label{belowN}
\end{figure}

Figure \ref{belowN} shows a series of traces obtained at a base temperature of $T=5$ K for various time delays between the pump ($1.55$ eV) and push ($0.6$ eV) pulses organized such that the push pulse is placed at the zero of the $t$ axis. The push pulse fluence here is $0.1$ mJ/cm$^2$. By comparing the behavior of the signal before and after the push pulse one can see that although the push pulse is causing visible kinks, the relaxation rate of the signal is not affected. Since the effect of the push pulse is only to increase the temperature of the medium, this observation is consistent with the general behavior of Na$_2$IrO$_3$ samples reported in \cite{alpichshev}. Indeed, provided the power of the push pulse is not enough to ``melt'' the order by increasing the local temperature above T$_N$, the relaxation time constant will not be affected.

A significantly different behavior can be observed for more intense push pulses. In Fig.\ref{aboveN}a we show a set of traces also obtained at $T=5$ K for push pulse fluence of $0.5$ mJ/cm$^2$ organized similarly to Fig.\ref{belowN}. The first thing to notice is that unlike Fig.\ref{belowN} there is a pronounced qualitative change in the time dependence caused by the push pulse. In particular, comparing the after-push (right) segment of the traces with the transient pump probe responses above $T_N$ in \cite{alpichshev} one can estimate that the push pulse is heating the system to about $T\approx 30\,K$. Importantly, a close inspection of the behavior of the signal right after the push pulse reveals that the immediate effect of the push pulse amounts only to a vertical shift of the signal for all traces independent of the pump-push delay [notice that $\Delta_1=\Delta_2$ in Fig. \ref{aboveN}(b)] which should be attributed to the production of additional ``parasitic'' electronic excitations by the push pulse. The creation of these excitations is most probably due to the limitations of the localized $J_{\text{eff}}=1/2$ moment picture of magnetism in Na$_2$IrO$_3$ by the push pulse \cite{choi}. They would be impossible had the local moments picture been precise \cite{foyevtsova}. But aside from the overall shift the fact that the difference between different traces remains unchanged across the push pulse strongly suggests that the configuration of pump-induced electronic excitations that was present before the arrival of the push pulse is not altered by it. Thus we conclude that the push pulse meets our requirements as a perturbation mainly causing an instantaneous increase in temperature while minimally interfering with the electronic configuration prior to it.

In light of the above, the data presented in Fig.\ref{aboveN} can be interpreted as follows: the pump pulse is creating doublons and holons which quickly form bound states. The subsequent dynamics can be viewed as the relaxation of HEs as a whole (the internal kinetic energy is dissipated rapidly on a few picosecond timescale), gradually releasing the excess of their kinetic energy via emission of magnetic excitations. When the push pulse arrives it melts the magnetic order but keeps the pre-push electronic configuration (including the non-equilibrium states of HEs) intact. The doublon-holon bound states are also known to exist above $T_N$ \cite{rixs, alpichshev}, therefore it is reasonable to think that the excitons ``survive'' the push pulse maintaining their kinetic energy, except that their structure changes from a bound state with a spin string [left inset to Fig. \ref{aboveN}(a)] to conventional bound state held together by Coulomb potential [right inset to Fig. \ref{aboveN}(a)].

Figure \ref{aboveN}(c) shows the main finding of this work. The top (blue) curve in this logarithmic plot is the time  dependence of the signal before push pulse showing the dynamics of the HE in the ordered phase. The lower (red) curve is the relaxation of the exciton in the push-induced disordered state produced after the push pulse. To produce the second curve we take the difference between two traces with different pump-push delays and thus corresponding to Hubbard excitons at different stage of relaxation [we use traces from Fig. \ref{aboveN}(b)]. By doing a subtraction we are getting rid of irrelevant components in the signal, including the contribution of the parasitic electrons which does not depend on the state of prepush excitons, and retrieve the information on the HE relaxation. Note that both curves are linear in a semilog plot and therefore are compatible with the simple relaxation picture adopted below. The conclusion from this figure is that a bound state without a spin string is relaxing much faster than the one with it.

The showings of Fig. \ref{aboveN}(c) can be interpreted by noting that the rate of relaxation of a nonequilibrium quasiparticle moving in a Mott insulator is directly proportional to its hopping integral $t_\text{eff}$. Indeed, every hopping process is associated with an emission of a spin excitation \cite{kane,lee}, therefore the more often the particle hops in a unit time (proportional to $t_\text{eff}$) the more quickly it loses its excessive energy. Since, as mentioned above the presence of a spin string significantly reduces the effective hopping integral of HE, the relaxation rate of an exciton in the ordered state is very slow \cite{zaanen}. In contrast eliminating the magnetic order makes the exciton lighter and the relaxation towards the quasi-equilibrium state happens much faster. An alternative way to look at this is to note that the presence of the string puts additional restrictions on the possible motions of doublons and holons therefore hindering the relaxation process. This shows that the enhanced total mass of a Hubbard exciton in the ordered state is indeed coming as a result of interaction with the magnetic medium around it. 

In conclusion we used a novel three-pulse ``pump-push-probe'' technique to address the issue of the behavior of quasiparticles moving in a frustrated Mott insulator. Applying different perturbations preferentially coupled to electronic and magnetic degrees of freedom in a time resolved manner we were able to trace the relaxation of the kinetic energy of the Hubbard excitons as a whole in both the magnetically ordered and disordered phases. We stress that this is fundamentally different from doing regular pump probe measurements at different static temperatures. Typically the relaxation of a correlated system is a complicated process and tracing the contribution of different degrees of freedom is often impossible. Here we are able to intervene in the process at any stage of development and use this knowledge to extract the details of the relevant sub-process. We observe that in the ordered phase the effective mass of Hubbard excitons is much larger as compared to the disordered state due to the presence of a spin string in the first case. This provides direct experimental evidence of the theoretical notion that the mass of a quasiparticle in a frustrated Mott insulator has a predominantly ``magnetic'' origin. Interestingly, there is a parallel phenomenon in high-energy physics, namely the fact that the majority of the mass of hadrons is coming not from the masses of constituent quarks but from the energy of the gluon field holding them together \cite{durr}. This is especially curious given that the spin-string mediated attraction between the doublon and holon is a direct analog of the quark confinement in QCD \cite{polyakov, senthil}.     

ZA gratefully acknowledges discussions with P. A. Lee and A. Kemper. A conversation with J. Zaanen was instrumental in clarifying the physical picture described in this paper. We would also like to thank A. Kogar for thoroughly reading the manuscript and making valuable comments. This work was supported by Army Research Office Grant No. W911NF-15-1-0128 and Gordon and Betty Moore Foundation EPiQS Initiative through Grant No. GBMF4540 (time resolved optical spectroscopy), Skoltech, as part of the Skoltech NGP program (theory) and National Science Foundation Grant No. DMR-1265162 (material growth).


\begin{thebibliography}{99}

\bibitem{pines}
D. Pines, P. Nozieres, \textit{The Theory Of Quantum Liquids} (W. A. Benjamin, New York, 1966)), Vol.1 .

\bibitem{imada}
M. Imada, A. Fujimori, Y. Tokura, Rev. Mod. Phys. 70, 1039 (1998).

\bibitem{lee} 
P. A. Lee, N. Nagaosa, X.-G. Wen, Rev. Mod. Phys. 78, 17 (2006).

\bibitem{brinkman}
W.F.Brinkman, T.M.Rice, Phys.Rev.B 2, 1324 (1970).

\bibitem{fradkin}
E. Fradkin, \textit{Field Theories of Condensed Matter Physics}, 2nd ed. (Cambridge University Press, Cambridge, England, 2013).

\bibitem{rokhsar}
S. A. Kivelson, D. S. Rokhsar, $\&$ J. P. Sethna, Phys. Rev. B 35 (16), 8865-8868 (1987).

\bibitem{khomski}
L.N.Bulaevskii, E.L.Nagaev, D.I.Khomskii, JETP 27, 836 (1968).

\bibitem{shraiman}
B. I. Shraiman, E. D. Siggia, Phys. Rev. Lett. 60, 740 (1988).

\bibitem{trugman}
S.A.Trugman, Phys. Rev. B 37, 1597 (1988).

\bibitem{varma}
S. Schmitt-Rink, C. M. Varma, A. E. Ruckenstein, Phys. Rev. Lett. 60, 2793 (1988).

\bibitem{kane}
C. L. Kane, P. A. Lee, and N. Read, Phys. Rev. B 39, 6880 (1989).

\bibitem{kim}
C. Kim, A. Y. Matsuura, Z.-X. Shen, N. Motoyama, H. Eisaki, S. Uchida, T. Tohyama, and S. Maekawa, Phys. Rev. Lett. 77, 4054 (1996).

\bibitem{thomas}
G. A. Thomas, J. Orenstein, D. H. Rapkine, M. Capizzi, A. J. Millis, R. N. Bhatt, L. F. Schneemeyer, J. V. Waszczak, Phys. Rev. Lett. 61, 1313 (1988).

\bibitem{peskin}
M. E. Peskin and D. V. Schroeder, \textit{An Introduction to Quantum Field Theory} (Westview Press, Boulder, Colorado, 1995).

\bibitem{durr}
S. D{\" u}rr, Z. Fodor, J. Frison, C. Hoelbling, R. Hoffmann, S. D. Katz, S. Krieg, T. Kurth, L. Lellouch, T. Lippert, K. K. Szabo, G. Vulvert, Science 322, 1224 (2008).

\bibitem{pekar}
L. D. Landau and S. I. Pekar, Zh. Eksp. Teor. Fiz. 18, 419 (1948).

\bibitem{misch1}
A. S. Mishchenko, N. Nagaosa, Phys. Rev. Lett. 93, 036402 (2004).

\bibitem{misch2}
A. S. Mishchenko, N. Nagaosa, K. M. Shen, Z.-X. Shen, X. J. Zhou and T. P. Devereaux, Europhys. Lett. 95, 57007 (2011).

\bibitem{kittel}
C. Kittel, \textit{Introduction to Solid State Physics} (John Wiley \& Sons, New York, 2005).

\bibitem{khaliullin1}
G. Jackeli and G. Khaliullin, Phys. Rev. Lett. 102, 017205 (2009).

\bibitem{khaliullin2}
J. Chaloupka, G. Jackeli, and G. Khaliullin, Phys. Rev. Lett. 105, 027204 (2010).

\bibitem{kitaev}
A. Kitaev, Ann. Phys. (N.Y.) 321, 2 (2006).


\bibitem{alpichshev}
Zh. Alpichshev, F. Mahmood, G. Cao, N. Gedik, Phys. Rev. Lett. 114, 017203 (2015).

\bibitem{hinton}
J. P. Hinton, S. Patankar, E. Thewalt, A. Ruiz, G. Lopez, N. Breznay, A. Vishwanath, J. Analytis, J. Orenstein, J. D. Koralek, I. Kimchi, Phys. Rev. B 92, 115154 (2015).

\bibitem{nembrini}
N. Nembrini, S. Peli, F. Banfi, G. Ferrini, Y. Singh, P. Gegenwart, R. Comin, K. Foyevtsova, A. Damascelli, A. Avella, C. Giannetti, arXiv:1606.01667 (2016).

\bibitem{rixs}
H. Gretarsson, J. P. Clancy, X. Liu, J. P. Hill, E. Bozin, Y. Singh, S. Manni, P. Gegenwart, J. Kim, A. H. Said, D. Casa, T. Gog, M. H. Upton, H. S. Kim, J. Yu, V. M. Katukuri, L. Hozoi, J. van den Brink, Y. J. Kim, Phys. Rev. Lett. 110, 076402 (2013).

\bibitem{orenstein}
J. Orenstein, Physics Today 65(9), 44 (2012).

\bibitem{demler}
R. Sensarma, D. Pekker, E. Altman, E. Demler, N. Strohmaier, D. Greif, R. Jordens, L. Tarruell, H. Moritz, T. Esslinger, Phys. Rev. B 82, 224302 (2010).

\bibitem{choi}
S. K. Choi, R. Coldea, A. N. Kolmogorov, T. Lancaster, I. I. Mazin, S. J. Blundell, P. G. Radaelli, Y. Singh, P. Gegenwart, K. R. Choi, S. W. Cheong, P. J. Baker, C. Stock, J. Taylor, Phys. Rev. Lett. 108, 127204 (2012).

\bibitem{damascelli}
R. Comin, G. Levy, B. Ludbrook, Z. H. Zhu, C. N. Veenstra, J. A. Rosen, Y. Singh, P. Gegenwart, D. Stricker, J. N. Hancock, D. van der Marel, I. S. Elfimov, A. Damascelli, Phys. Rev. Lett. 109, 266406 (2012).

\bibitem{bjkim1}
B. J. Kim, H. Jin, S. J. Moon, J. Y. Kim, B. G. Park, C. S. Leem, J. Yu, T. W. Noh, C. Kim, S. J. Oh, J. H. Park, V. Durairaj, G. Cao, E. Rotenberg, Phys. Rev. Lett. 101, 076402 (2008).

\bibitem{rupprecht}
G. Rupprecht, Phys. Rev. Lett. 12, 580 (1964).

\bibitem{bloembergen}
Y. R. Shen and N. Bloembergen, Phys. Rev. 137, A1787 (1965).

\bibitem{fleury}
P. A. Fleury and R. Loudon, Phys. Rev. 166, 514 (1968).

\bibitem{herring}
R. M. White, R. J. Nemanich, C. Herring, Phys. Rev. B 25, 1822 (1982).

\bibitem{raman}
S. N. Gupta1, P. V. Sriluckshmy, K. Mehlawat, A. Balodhi, D. K. Mishra, S. R. Hassan, T. V. Ramakrishnan, D. V. S. Muthu, Y. Singh and A. K. Sood, Europhys. Lett. 114, 47004 (2016).

\bibitem{supp}
See Supplemental Material at http://link.aps.org/supplemental/10.1103/PhysRevB.xx.xxxxxx for the discussion of the effects of the push pulse.

\bibitem{sample1}
M. Ge, T. F. Qi, O. B. Korneta, D. E. De Long, P. Schlottmann, W. P. Crummett, G. Cao, Phys. Rev. B 84, 100402(R) (2011).

\bibitem{sample2}
S. Chikara, O. Korneta, W. P. Crummett, L. E. De Long, P. Schlottmann, G. Cao, Phys. Rev. B 80, 140407(R) (2009).

\bibitem{sample3}
M. A. Laguna-Marco, D. Haskel, N. Souza-Neto, J. C. Lang, V. V. Krishnamurthy, S. Chikara, G. Cao, and M. van Veenendaal, Phys. Rev. Lett. 105, 216407 (2010).

\bibitem{foyevtsova}
I. I. Mazin, H. O. Jeschke, K. Foyevtsova, R. Valenti, and D. I. Khomskii, Phys. Rev. Lett. 109, 197201 (2012).

\bibitem{zaanen}
L. Rademaker, K. Wu, H. Hilgenkamp, J. Zaanen, Europhys. Lett. 97, 27004 (2012).

\bibitem{polyakov}
A. M. Polyakov, \textit{Gauge Fields and Strings} (Harwood Academic, Chur, Switzerland, 1987).

\bibitem{senthil}
T. Senthil, M. P. A. Fisher, J. Phys. A: Math. Gen. 34, L119 (2001).


\end{thebibliography}
\end{document}